\documentclass[aps,pra,showpacs,superscriptaddress,amssymb,twocolumn]{revtex4-1}
\usepackage{graphicx}
\usepackage{appendix} 
\usepackage{hyperref}
\usepackage{color}\hypersetup{colorlinks=true,citecolor=blue,linkcolor=black}
\usepackage{bm,amsmath}
\usepackage{dcolumn}
\usepackage{amsthm}

\newcommand{\be}{\begin{equation}}
\newcommand{\ee}{\end{equation}}

\begin{document}

\rightline{\tt }

\vspace{0.2in}

\title{Efficient correction of multiqubit measurement errors}

\author{Michael R. Geller}
\author{Mingyu Sun}
\email{mingyu.sun25@uga.edu}
\affiliation{Center for Simulational Physics, University of Georgia, Athens, Georgia 30602, USA}
\date{\today}

\begin{abstract}
State preparation and measurement (SPAM) errors limit the performance of near-term quantum computers and their potential for practical application. SPAM errors are partly correctable after a calibration step that requires, for a complete implementation on a register of $n$ qubits, $2^n$ additional measurements. Here we introduce an approximate but efficient method for multiqubit SPAM error characterization and mitigation requiring the classical processing of $2^n \! \times 2^n$ matrices, but only $O(4^k n^2)$ measurements, where $k=O(1)$ is the number of qubits in a correlation volume. We demonstrate and validate the technique using an IBM Q processor on registers of 4 and 8 superconducting qubits.
\end{abstract}

\maketitle

\section{Introduction}

Errors in a quantum computation are typically classified into state-preparation errors, gate errors, and measurement errors. Quantum error correction has mostly focused on gate errors, in part because state preparation and measurement (SPAM) errors only occur at the beginning and end of a circuit or error-correction cycle, and, unlike gate errors, do not accumulate with circuit depth \cite{Lidar2013}. A consequence of this difference is that surface code error correction, a practical route to fault-tolerant quantum computation, is significantly less sensitive to measurement errors than to gate errors, tolerating errors many times larger \cite{FowlerPRA12}.

However in near-term quantum processors, which are not error corrected, measurement errors are large and, even worse, may increase with register size due to increased crosstalk. A common approach for correcting some of the SPAM error is to measure the transition matrix $T$ between all initial and final classical states, and then use this information to classically correct subsequently measured data \cite{BialczakNatPhys10,NeeleyNat10,DewesPRL12,14114994,160304512,SongPRL17,181102292,190505720,180411326,TannuIEEE19,190411935,190503150,190708518,191000129,191001969,191113289,200104449,200601805}.  We refer to this error mitigation technique as $T$ matrix SPAM correction, as it attempts to correct measurement errors (also known as readout or assignment errors) as well as the smaller state preparation errors.

\subsection{T matrix SPAM correction}

Let the set of $n$ physical qubits we wish to correct be called the {\it register} (the register does not have to include every qubit in the processor). The $T$ matrix SPAM correction technique can be described as follows:  Let $x, x'  \in \lbrace 0,1\rbrace^n$ be classical states of the $n$ qubits in the register, and define elements of a $2^n \! \times  \! 2^n$ matrix $T$ by
\begin{equation}
\! T(x|x') \! = \! T(x_1 \cdots x_n|x'_1  \cdots x'_n) \! = \!  {\rm Tr} [
E_{x_1 \cdots x_n} \rho_{x'_1  \cdots x'_n}].
\label{defT}
\end{equation}
Here $E_{x}$ is the multiqubit POVM element characterizing the nonideal implementation of the projector $|x\rangle \langle x |$, and $ \rho_{x'}$ is the density matrix produced after attempting to prepare classical state $|x'\rangle \langle x' |$. Each column $x'$ of $T$ is the raw probability distribution ${\rm prob}(x),$ measured immediately after preparing $x'$. 
In the absence of any SPAM error,
\begin{equation}
T(x|x')=\delta_{x x'}, \ \ \ {\rm (ideal)}
\end{equation}
 the $2^n \! \times  \! 2^n$ identity. The complete implementation of the technique is to measure $T$ and classically apply $T^{-1}$ to subsequently measured probability distributions \cite{BialczakNatPhys10,NeeleyNat10,DewesPRL12,14114994,160304512,SongPRL17,181102292,190505720,180411326,TannuIEEE19,190411935,190503150,190708518,191000129,191001969,191113289,200104449,200601805}. This forces an empty circuit in the noisy processor to act ideally. 

However there are several limitations of  this approach:  (i) The complete implementation requires $2^n$ characterization experiments (probability measurements), which is not scalable. The classical processing of the calibration data is also inefficient. (ii) The matrix $T$ may become singular for large $n$, preventing direct inversion. (iii) The inverse $T^{-1}$ might not be a stochastic matrix, meaning that it can produce negative corrected probabilities. (iv) The correction is not rigorously justified, so we cannot be sure that we are only removing SPAM errors and not otherwise corrupting an estimated probability distribution.

Limitations (i) and (ii) can be circumvented \cite{SongPRL17,181102292,190505720} by approximating $T$ with a tensor product of single-qubit transition matrices $T_i$,
\begin{eqnarray}
T_{\rm prod} &=&  T_1 \otimes T_2  \otimes \cdots  \otimes T_n , 
\label{Tprod} \\
T_{\rm prod}^{-1} &=&  T_1^{-1}  \otimes T_2^{-1}   \otimes \cdots  \otimes T_n^{-1} . 
\label{T_1prod}
\end{eqnarray}
Each $T_i$ is a $2 \times  2$ matrix with elements  $T_i(x_i|x_i')$, measured by initializing qubit $i$ with classical state $x_i' \in \{ 0,1 \}$ and measuring ${\rm prob}(x_i)$. We show here that the accuracy of (\ref{Tprod}) is rather poor, however, especially on registers with significant crosstalk. The low accuracy indicates the presence of large multiqubit correlations, which we identify below.

Limitation (iii) is easily avoided by minimizing
\begin{equation}
\| T \, p_{\rm corr} - p_{\rm raw} \|^2_2 
\end{equation}
subject to physicality constraints $0 \le p_{\rm corr}(x) \le 1$ and $\| p_{\rm corr} \|_1 =1$. Alternatively, one can use constrained maximum-liklihood estimation \cite{190411935} or iterative Bayesian unfolding \cite{191001969} here. Limitation (iv) was addressed recently in Refs.~\cite{190708518} and \cite{200201471}, where it was shown that a certain family of nonideal quantum measurements (having diagonal POVMs in the classical basis) can be perfectly corrected, at least in principal.

\subsection{This work}

In this work, we go beyond the product approximation (\ref{Tprod}) by deriving an efficient yet accurate method to estimate $T$. The technique is efficient in the sense that it only requires $O(n^2)$ probability measurements to estimate the entire set of $4^n$ matrix elements $\lbrace T(x|x') \rbrace_{x,x'}$. However evaluating these $4^n$ elements from the measured data remains classically inefficient. While $T$ matrix SPAM correction might ultimately be unscalable, we envision that the technique introduced here will enable error mitigation on large registers of qubits, greatly extending the reach and power of near-term quantum computing. We note that fully scalable SPAM correction is included in any fault-tolerant quantum computing framework, at the expense of significant qubit overhead \cite{Lidar2013}.

The organization of our paper is as follows: Section \ref{Methods} discusses the online superconducting qubits, Moore neighborhoods, and qubit filters used in this work, as well as the error measures used to report the results. Section \ref{Total SPAM error} discusses measures of the overall magnitude of multiqubit SPAM errors. Section~\ref{Product approximation} assesses the accuracy of the product approximation (\ref{Tprod}). Section~\ref{Multiqubit measurement error correlators} introduces and measures the multiqubit correlation functions  used for scalable $T$ matrix estimation. Section \ref{Scalable T matrix estimation} describes and implements the scalable estimation technique. Section~\ref{Conclusions} contains our conclusions.

\section{Methods}
\label{Methods}

In this section we discuss the online superconducting qubits used in this work, and we introduce two theoretical tools, Moore neighborhoods and qubit filters, used below. Finally, we discuss the error measures used to report our experimental results.

\subsection{Qubits}

\begin{figure}
\includegraphics[width=7.5cm]{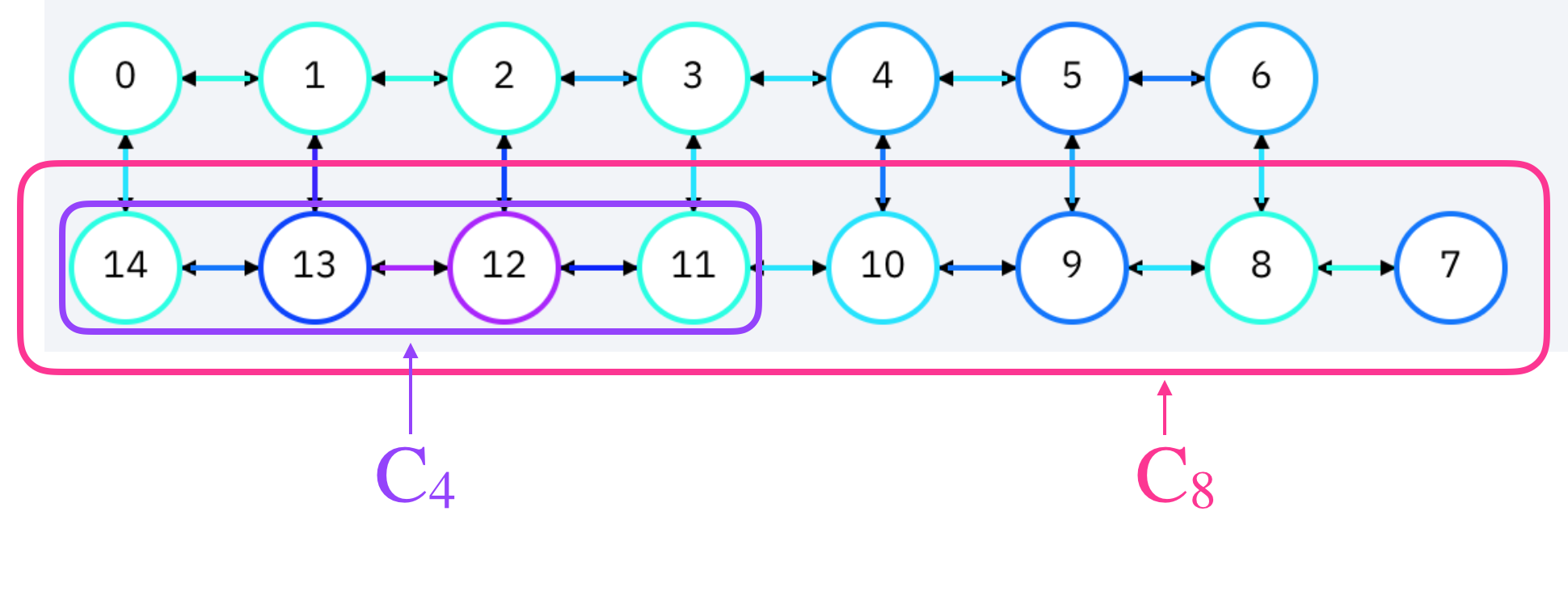} 
\caption{Layout of IBM Q device ibmq\_16\_melbourne. In this work we use the registers $C_4$ and $C_8$.}
\label{melborneChains}
\end{figure}

The experiments reported here were performed with the IBM Q processor ibmq\_16\_melbourne, based on superconducting transmon qubits, and the BQP online data acquisition software developed by the authors.  BQP is a Python package developed to design, run, and analyze complex quantum computing and quantum information experiments using commercial backends. 
In this work we validate the SPAM correction technique on two registers of the ibmq\_16\_melbourne device, shown in Fig.~\ref{melborneChains}. We choose linear chains 
\begin{equation}
C_4 = \{ Q_{14}, Q_{13}, Q_{12}, Q_{11}  \}
\label{defC4}
\end{equation}
and
\begin{equation}
C_8 = \{ Q_{14}, Q_{13}, Q_{12}, Q_{11}, Q_{10}, Q_{9}, Q_{8}, Q_{7} \} 
\label{defC8}
\end{equation}
to enable a separation of short- and long-range correlations. And we intentionally overlap the registers to highlight the register-dependence of individual single-qubit SPAM errors.
 
\subsection{Moore neighborhoods}
\label{Moore neighborhoods}

In this work we incorporate locality through the use of qubit neighborhoods. For a given register of $n$ physical qubits with positions ${\vec r}_i \in \mathbb{R}^D$, where $i \in \{ 1,2, \cdots ,n\}$, let ${\cal N}_i$ be the set of $k$ qubits closest to $i$ in Chebyshev distance $\|{\vec r}_j- {\vec r}_i \|_{\infty}$.  Here 
\begin{equation}
\| {\vec r} \|_{\infty} = \lim_{p \rightarrow \infty} \bigg( \sum_{i=1}^D |r_i|^p \bigg)^{\! 1/p} \! \!
= \max \, \{  |r_1|,\cdots, |r_D|  \}.
\end{equation}
We call ${\cal N}_i$ the {\it neighborhood} of qubit or site $i$, noting that ${\cal N}_i$ does not include qubit $i$.  The nonnegative integer $k$ is the {\it size} of the neighborhood and is independent of $i$. A formal definition of ${\cal N}_i$ (applying to any array of qubits) will not be provided here. Instead we provide a definition of ${\cal N}_i$ for $D$-dimensional square lattices, and argue that any reasonable extension to other arrays, appropriately incorporating qubit locality, will be sufficient. 

For infinite square latices with spatial dimension $D=1,2,3,\cdots$, the neighborhood ${\cal N}_i$ is uniquely defined for sizes
\begin{equation}
k \in \bigg\lbrace (2 \ell + 1)^D - 1, \ \ \ell =0,1,2,3,\cdots \bigg\rbrace .
\label{ks}
\end{equation}
The corresponding ${\cal N}_i$ are called {\it Moore neighborhoods} with range $\ell$ \cite{mooreNote}. In our application, $\ell$ is the number of layers in ${\cal N}_i$ surrounding qubit $i$, which is itself excluded. $k=0$ corresponds to empty neighborhoods, relevant when there are no multiqubit correlations. 

\begin{figure}
\includegraphics[width=7cm]{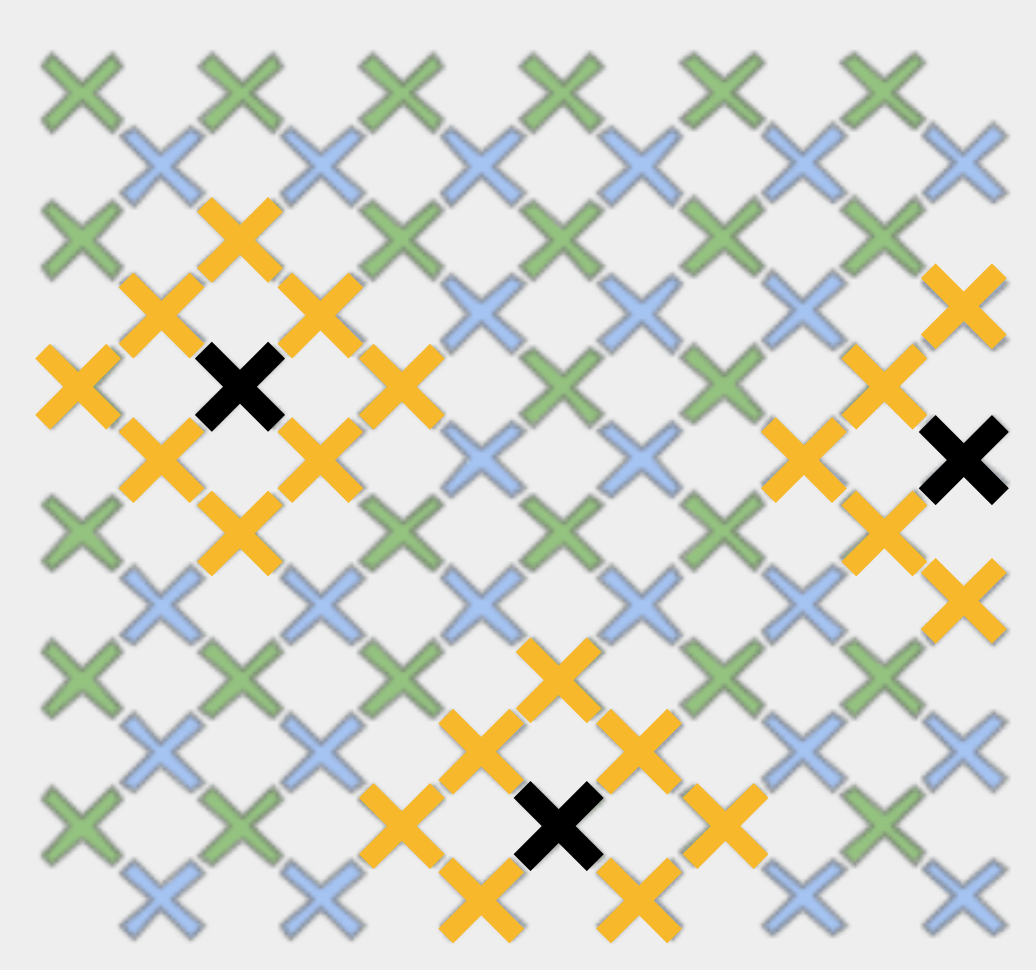} 
\caption{$k\!=\!8$ Moore neighborhoods (yellow crosses) on Google's Bristlecone chip.}
\label{bristleconeNeighborhood}
\end{figure}

Actual qubit arrays are finite, of course, and have spatial boundaries. We extend the definition of the Moore neighborhoods in this case as follows: We restrict $k$ to a bulk value from the set (\ref{ks}). Then, if qubit $i$ is near a boundary, we truncate ${\cal N}_i$, as defined for an infinite lattice, by excluding the missing qubits. A 2d example is illustrated in Fig.~\ref{bristleconeNeighborhood}. Therefore, for $i$ near a boundary, ${\cal N}_i$ may contain fewer than $k$ qubits.

\subsection{Qubit filters}
\label{Qubit filters}

We use qubit filters to simplify and reduce the number of circuits that have to be measured for $T$ estimation. For a given register of qubits and their size-$k$ Moore neighborhoods ${\cal N}_i,$ we define functions $f_{i}(x)$ and $f_{ij}(x)$ called {\it filters} that
act on classical states $x = x_1 \cdots x_n$ according to
\begin{equation}
f_i(x_1 \cdots x_n) = y_1 \cdots y_n \ {\rm with} \  y_l = 
\begin{cases} x_i \ {\rm for} \ l=i  \\  
x_i \ {\rm for}\ l \in {\cal N}_i \\
0 \ {\rm else}  \end{cases}
\end{equation}
and
\begin{equation}
f_{ij}(x_1 \cdots x_n) = y_1 \cdots y_n \ {\rm with} \  y_l = 
\begin{cases} x_i \ {\rm for} \ l=i, j  \\  
x_i \ {\rm for} \ l \in {\cal N}_i \cup {\cal N}_j \\
0 \ {\rm else}  \end{cases} \! \! \! \! .
\end{equation}
For 1d registers these simplify to
\begin{equation}
f_{i}(x) = 0 \cdots 0 x_{i-\frac{k}{2}} \cdots  x_i \cdots x_{i+\frac{k}{2}} 0 \cdots  0
\end{equation}
and
\begin{equation}
f_{ij}(x) = 0 \cdots 0 x_{i-\frac{k}{2}} \cdots x_{i+\frac{k}{2}} 0 \cdots  0
x_{j-\frac{k}{2}} \cdots x_{j+\frac{k}{2}} 0 \cdots 0.
\end{equation}
The filters have the effect of setting bit values far away from site $i$ (or sites $i,j$) to 0. 

\subsection{Error measures}
\label{Error measures}

Quantifying multiqubit SPAM requires physically meaningful error measures. A standard measure of single-qubit SPAM error is \cite{14114994}
\begin{equation}
\epsilon = \frac{T(0|1) + T(1|0)}{2}.
\label{defEpsilonError}
\end{equation}
$T(0|1)$ is the probability of observing $|0\rangle$ when $|1\rangle$ is prepared. $T(1|0)$ is the opposite. 

Here we propose two norms appropriate for multiqubit $T$ matrices. One is a scaled Frobenius norm
\begin{equation}
\| A \|_d := \frac{ \| A \|_{\rm F}  }{\sqrt{  \dim(A) }} = \sqrt{ \frac{\sum_{xx'} |A_{xx'}|^2  }{ \dim(A) }} ,
\label{defScaledFrobenius}
\end{equation}
where $\dim(A)$ is the dimension of the matrix $A$. The scale factor in (\ref{defScaledFrobenius}) enables SPAM errors to be compared between qubit registers with different sizes. Specifically, the norm $\| \cdot \|_d $ has the asymptotic property that, for uncorrelated errors, it leads to an error measure that scales linearly with register size $n$, for small symmetric errors and large $n$.  Let
\begin{equation}
\tau = \begin{pmatrix} 1-\epsilon & \epsilon \\ \epsilon & 1-\epsilon \\ \end{pmatrix}\end{equation}
be a symmetric single-qubit $T$ matrix with SPAM error $\epsilon$. We prove in Appendix A that  
\begin{equation}
\lim_{n \rightarrow \infty} \lim_{\epsilon \rightarrow 0} \bigg \| \begin{pmatrix} 1-\epsilon & \epsilon \\ \epsilon & 1-\epsilon \\ \end{pmatrix}^{\! \! \otimes n} \! \!  - I \  \bigg \|_d = n \, \epsilon.
\end{equation}
Here $I$ is the $2^n \! \times 2^n$ identity representing the ideal value of the multiqubit $T$ matrix.

The second  norm we use is the max norm
\begin{equation}
\| A \|_{\rm max} = \max_{x x'} |A_{xx'} |, 
\end{equation}
which is useful for identifying the largest magnitude of error in the individual  matrix elements. 

\section{Total SPAM error}
\label{Total SPAM error}

To characterize the overall magnitude of multiqubit SPAM errors and develop an intuition for these norms, we measure the {\it total} SPAM error \cite{191113289}
\begin{equation}
\| T_{\rm meas} - I \|
\label{defTotalSPAMError}
\end{equation}
on the ibmq\_16\_melbourne registers $C_4$ and $C_8$. Here $T_{\rm meas}$ is the directly measured $2^n \! \times \! 2^n$ transition matrix. We also measure the single-qubit SPAM error (\ref{defEpsilonError}) for each qubit, and the register average. The individual errors are provided in the column labelled ``$\epsilon \, ({\rm meas})$" in Tables \ref{tableSingleQubitRegistersize4} and \ref{tableSingleQubitRegistersize8} of Appendix \ref{Single-qubit T matrices}.  The average values are summarized here in Table \ref{totalSPAMErrorTable}. 

\begin{table}[htb]
\centering
\caption{Total SPAM error measured on ibmq\_16\_melbourne registers $C_4$ and $C_8$ \cite{dataTakingMarch4}. Here $n$ is the register size and
$\epsilon$ is the measured single-qubit SPAM error (\ref{defEpsilonError}) averaged over the register. The final columns give the total SPAM error (\ref{defTotalSPAMError}) measured in two different norms. Each circuit was measured with 32k samples.}
\begin{tabular}{|c|c|c|c|c|}
\hline
$n$ & $\epsilon$  & $n \, \epsilon$ &   $\| T_{\rm meas} - I \|_d$  & $\| T_{\rm meas} - I \|_{\rm max}$  \\
\hline 
4 &  0.052  & 0.209 & 0.259  & 0.347 \\
8 & 0.086 & 0.686 & 0.571 & 0.668   \\
\hline
\end{tabular}
\label{totalSPAMErrorTable}
\end{table}

The total SPAM errors reported in Table \ref{totalSPAMErrorTable} are huge. On the 4-qubit chain, the largest single error in $T_{\rm meas}$ is $34.7\%$. On the 8-qubit chain, the largest error is $66.8\%$. These large values make accurate SPAM mitigation highly challenging!

\section{Product approximation}
\label{Product approximation}

In this section we discuss the approximation (\ref{Tprod}), which neglects multiqubit error correlation, and measure its accuracy on the ibmq\_16\_melbourne registers $C_4$ and $C_8$ shown in Fig.~\ref{melborneChains}. The accuracy is deduced by measuring the $n$-qubit $T$ matrix, the single-qubit matrices  $T_i$ for all qubits in the register, and calculating $\| T_{\rm prod}  - T_{\rm meas} \|$ in the  norms defined in Sec.~\ref{Error measures}.

\subsection{Single qubit T matrices}
\label{Single qubit T matrices}

First we will need to define a single-qubit matrix $T_i$ for qubit $i$. Let the other qubits $j \neq i$ in the register be called {\it spectator} qubits. There are several natural definitions of $T_i$ in a multiqubit register, differing by how the spectator qubits are initialized. In this section we consider two families, called {\it uniform} and {\it average}, and assess the accuracy of the product approximation using both variants.

In the uniform family, all spectator qubits are initialized to the same value, 0 or 1. The columns of the $T_i$ are obtained by measuring ${\rm prob}(x_1 x_2 \cdots x_n)$ and tracing over the spectator qubits, 
\begin{equation}
{\rm prob}(x_i) = \bigg( \prod_{j \neq i} \sum_{x_j=0}^1 \bigg) {\rm prob}(x_1 x_2 \cdots x_n).
\end{equation}
The single-qubit $T$ matrices for this family are given in the columns labelled 
\begin{eqnarray}
T_i \, ({\rm spectators} \! = \!  |0\rangle, |1\rangle) 
\end{eqnarray}
in Tables \ref{tableSingleQubitRegistersize4} and 
\ref{tableSingleQubitRegistersize8}
of Appendix \ref{Single-qubit T matrices}. 

In the other family, we average the $T$ matrix over the initial conditions of   $k$ neighboring spectator qubits. That is, for each qubit $i$, we consider the set of (up to) $k$ spectator qubits in the Moore neighborhood ${\cal N}_i$. The single-qubit transition matrix for qubit $i$ is measured for each of the $2^{k}$ initial conditions of the spectators and then averaged. Spectator qubits outside of the neighborhood are initialized to 0. The single-qubit $T$ matrices for this family are given in the columns labelled 
\begin{equation}
T_i \ ({\rm ave} \! :  \,  k \! = \! 2,4,6) 
\end{equation}
in Tables \ref{tableSingleQubitRegistersize4} and 
\ref{tableSingleQubitRegistersize8}
of Appendix \ref{Single-qubit T matrices}. Below we find that this definition of single-qubit $T$ matrix leads to the most accurate $T_{\rm prod}$.

\subsection{Results}
\label{Single qubit T matrices Results}

The accuracy of the tensor product approximation (\ref{Tprod}) on the two ibmq\_16\_melbourne registers is given in Tables \ref{tableTprodErrorRegistersize4} and \ref{tableTprodErrorRegistersize8}.
We see that the errors $ \| T_{\rm prod}-T_{\rm meas} \|$ in the product approximation are considerably smaller than the total SPAM errors given in Table \ref{totalSPAMErrorTable}. This might suggest that the product approximation, which neglects multiqubit error correlation, is acceptable. However the error in the individual elements (see max norms) can be as large as $\sim \! \! 5\%$, which is prohibitive for many applications.

\begin{widetext}

\begin{table}[htb]
\centering
\caption{Accuracy of $T_{\rm prod}$ on the register $C_4$ shown in Fig.~\ref{melborneChains}. Here $n$ is the register size. Each circuit was measured with 32k samples.}
\begin{tabular}{|c|}
\hline
\\
$n=4$  \\
 \\
\hline 
scaled Frobenius \\
\hline
max norm \\
\hline
\end{tabular}
\begin{tabular}{|c|}
\hline
\\
$ \| T_{\rm prod}-T_{\rm meas} \| $ \\
\hline 
\begin{tabular}{c|c|c}
$T_i \ ({\rm spectators} \! = \!  |0\rangle)$ & $T_i \ ({\rm spectators} \! = \!  |1\rangle)$ & $T_i \ ({\rm ave} \! :  \,  k \! = \! 2) $  \\
\hline 
0.044 & 0.048 & 0.044  \\
\hline 
0.055  & 0.053 & 0.053  \\
\hline
\end{tabular}
\\
\hline
\end{tabular}
\label{tableTprodErrorRegistersize4}
\end{table}

\begin{table}[htb]
\centering
\caption{Accuracy of $T_{\rm prod}$ on the register $C_8$ shown in Fig.~\ref{melborneChains}. Each circuit was measured with 32k samples.}
\begin{tabular}{|c|}
\hline
\\
$n=8$  \\
 \\
\hline 
scaled Frobenius \\
\hline
max norm \\
\hline
\end{tabular}
\begin{tabular}{|c|}
\hline
\\
$ \| T_{\rm prod}-T_{\rm meas} \| $ \\
\hline 
\begin{tabular}{c|c|c|c|c}
$T_i \ ({\rm spectators} \! = \!  |0\rangle)$ & $T_i \ ({\rm spectators} \! = \!  |1\rangle)$ & $T_i \ ({\rm ave} \! :  \,  k \! = \! 2) $ & $T_i \ ({\rm ave} \! :  \,  k \! = \! 4) $ & $T_i \ ({\rm ave} \! :  \,  k \! = \! 6) $ \\
\hline 
0.048 & 0.050 & 0.046 & 0.045 & 0.036 \\
\hline 
0.067 & 0.081 & 0.069 & 0.068 & 0.056 \\
\hline
\end{tabular}
\\
\hline
\end{tabular}
\label{tableTprodErrorRegistersize8}
\end{table}

\end{widetext}
    
\section{Multiqubit measurement error correlators}
\label{Multiqubit measurement error correlators}

In Sec.~\ref{Product approximation} we demonstrated that the product approximation (\ref{Tprod}) has limited accuracy on the ibmq\_16\_melbourne registers, indicating the presence of significant multiqubit correlations. Here we confirm that conclusion by identifying and directly measuring those correlations. 

In this work we use three multiqubit correlation functions (called $A, B, C$)  to characterize SPAM errors.  These quantities result from adapting the language of mean fields and fluctuations to the transition matrix (\ref{defT}), where the product approximation (\ref{Tprod}) corresponds to an uncorrelated approximation for $T$, and the scalable estimation technique results from including quadratic and possibly higher-order fluctuations. 

A tensor product structure and locality of the measurement operators are essential to our approach. However these properties are not apparent in the definition (\ref{defT}), and are not rigorously present without an additional continuity assumption. We address this issue in the next section.
The correlators $A, B, C$ are defined in Sec.~\ref{T matrix and mean field theory}.

\subsection{Uncovering the tensor product structure}

Physically, the absence of an explicit tensor-product structure in the multiqubit $T$ matrix  (\ref{defT}) is caused by interactions between the physical qubits.
We use this fact to {\it undo} that interaction, thereby revealing the underlying tensor product structure. To each physical system or device we therefore introduce an associated {\it noninteracting} qubit array consisting of a register of $n$ qubits, each (optionally) coupled to its own independent measurement apparatus, but with no cross coupling between qubits or detectors. Our assumption is that states of the noninteracting and interacting systems are related by a completely positive trace-preserving (CPTP) map, 
\begin{equation}
\rho_{x'}^0 \mapsto \rho_{x'} = \Lambda (\rho_{x'_1}^{0} \otimes \cdots \otimes  \rho_{x'_n}^{0} ).
\label{CPTPmap}
\end{equation}
Here $\Lambda$ is a CPTP superoperator. The initial state  $\rho_{x'}^0$ of the noninteracting array is separable and can be written as a product $ \rho_{x'_1}^0 \otimes \cdots \otimes \rho_{x'_n}^0$ of single-qubit density matrices.  A sufficient condition for (\ref{CPTPmap}) to hold is that the states of the coupled qubits can be obtained by adiabatically turning on the qubit-qubit interaction $V$. In this case
\begin{equation}
\rho_{x'} =  S (\rho_{x'_1}^{0} \otimes \cdots \otimes  \rho_{x'_n}^{0} )S^\dagger,
\label{adiabaticMap}
\end{equation}
where $S = T e^{-i \int \! V \! dt}$ is a time-evolution operator.  Note that we don't require the map between the interacting and noninteracting limits to be adiabatic or even unitary. But we exclude cases where the initial state of the uncoupled register is entangled with the environment, and cases where turning on the qubit-qubit coupling causes leakage out of the register; in these cases the map would not be CPTP. 

The transition matrix in the noninteracting array is
\begin{equation}
{\rm Tr} \big[ E_{x_1}^{(1)} \otimes \cdots \otimes E_{x_n}^{(n)}  \, (\rho_{x'_1}^{0} \otimes \cdots \otimes  \rho_{x'_n}^{0} )\big].
\label{uncoupled}
\end{equation}
Here $E_{0}^{(i)}$ and $E_{1}^{(i)} = I - E_{0}^{(i)}$ are two-outcome POVM elements for qubit $i$, which may vary from qubit to qubit. Due to detector nonidealities, the $E_{x_i}^{(i)}$ may differ from projectors, but the multiqubit measurement operators are tensor products of the single-qubit ones as the qubits are uncoupled. Then using (\ref{CPTPmap}) we can write (\ref{defT}) as
\begin{eqnarray}
T(x|x') =
\big\langle E_{x_1}^{(1)} \otimes \cdots \otimes E_{x_n}^{(n)} \big\rangle_{\! x'},
\label{mapped}
\end{eqnarray}
where 
\begin{equation}
\langle O \rangle_{x'} =
{\rm Tr} [O \Lambda ( \rho_{x'_1}^0 \otimes \cdots \otimes  \rho_{x'_n}^0 ) ] 
\end{equation}
is the expectation value after preparing the noisy classical state $x' = x_1' x_2' \cdots x_n'$. 

In the remainder of this paper, we always refer to interacting qubit arrays. The map between noninteracting and interacting qubits simply allows us to assume a tensor product of measurement operators for the latter, at the expense of including the superoperator $\Lambda$.  We will not need the explicit form of $\Lambda$, however, because our approach ultimately derives {\it relations} between quantities in the interacting array, which get measured there.

\subsection{T matrix and mean field theory}
\label{T matrix and mean field theory}

In this section we explain the connection between $T$ matrices and mean field theory, and then define the multiqubit correlators $A, B, C$ used to characterize and mitigate SPAM errors. Recall that $E_{0}^{(i)}$ and $E_{1}^{(i)} \! =\! I  - E_{0}^{(i)}$ are nonideal single-qubit POVM elements for qubit $i$, which are $2 \! \times \! 2$ positive semidefinite Hermitian matrices tensored with identities on the spectator qubits (all qubits $j$ in the register other than $i$) . We can regard the $E_{x_i}^{(i)}$ as {\it operator fields} as a function of discrete qubit position $i$, which will become correlated in the presence of qubit-qubit coupling and crosstalk.

First we define mean fields
\begin{equation}
\langle E_{x_i}^{(i)} \rangle_{x'} =  {\rm Tr} [E_{x_i}^{(i)} \Lambda(\rho^0_{x'})],
\ \ \ x_i \in \{0,1\}
\label{meanField}
\end{equation}
for the measurement operators $E_{x_i}^{(i)} \! ,$ and also their fluctuations 
\begin{equation}
\delta E_{x_i}^{(i)} =  E_{x_i}^{(i)} -  \langle E_{x_i}^{(i)} \rangle_{x'} .
\label{measurementFluctuation}
\end{equation}
Here $x' \in \{0,1\}^n$ is a classical initial state and $\Lambda(\rho^0_{x'})$ is the noisy implementation of that state (in the interacting array). Note that our notation  (\ref{measurementFluctuation}) for the measurement fluctuation $\delta E_{x_i}^{(i)}$ suppresses its dependence on $x'$.

\begin{figure}
\includegraphics[width=8.5cm]{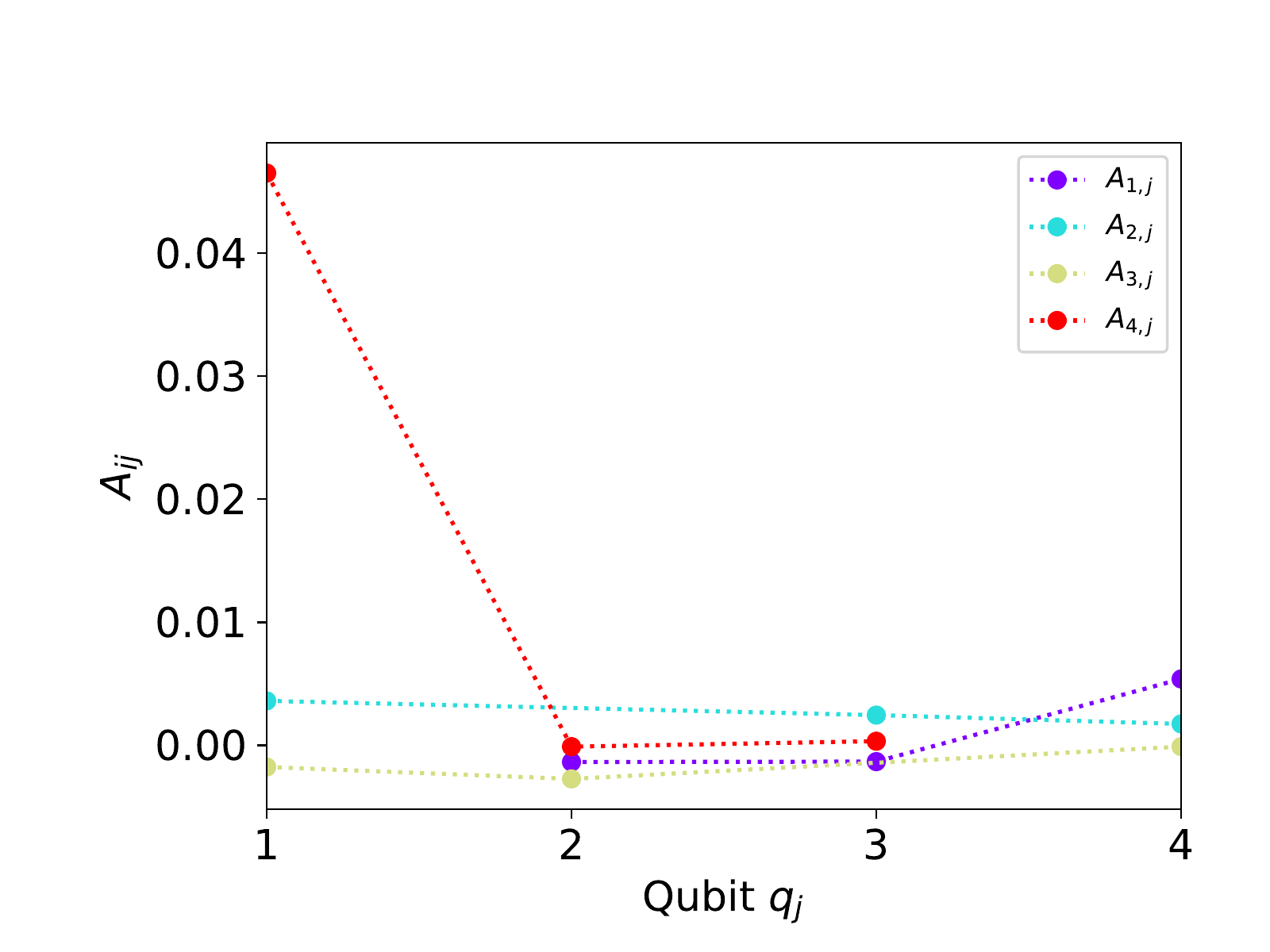} 
\caption{$A_{ij}$ correlator, defined in (\ref{defA}), for the 4-qubit register $C_4$. Each circuit was measured with 32k samples. In this figure, qubits $j=1,2,3,4$ refer to their position along the chain $Q_{14}, Q_{13}, Q_{12}, Q_{11}$. Note the large nonlocal spectator dependence $A_{41}=A_{Q_{11}, Q_{14}}=0.047$.}
\label{Acorrelator4}
\end{figure} 

The mean fields (\ref{meanField}) are closely related to the single-qubit transition matrices $T_i$ introduced in Sec.~\ref{Single qubit T matrices}, the precise relation depending on how spectator qubits are initialized. For $T_i$ in the uniform family with spectators set to $b \in \{ 0, 1 \}$, we have
\begin{eqnarray}
T_i(x_i |x_i' ) = \langle E_{x_i}^{(i)} \rangle_{b \cdots b x_i' b \cdots b} .
\label{defTiUniform}
\end{eqnarray}
For $T_i$ in the average family with $k$ neighbors, we have
\begin{eqnarray}
T_i(x_i |x_i' ) = \frac{1}{2^k} \sum_{y_1=0}^1 \cdots \sum_{y_k=0}^1   \langle E_{x_i}^{(i)} \rangle_{z} ,
\label{defTiAverage}
\end{eqnarray}
where 
\begin{equation}
z = z_1 \cdots z_n \ {\rm with} \  z_l = 
\begin{cases} x'_i \ {\rm for} \ l=i  \\  
y_l \ {\rm for}\ l \in {\cal N}_i \\
0 \ {\rm else}  \end{cases} \! .
\end{equation}

\begin{figure}
\includegraphics[width=8.5cm]{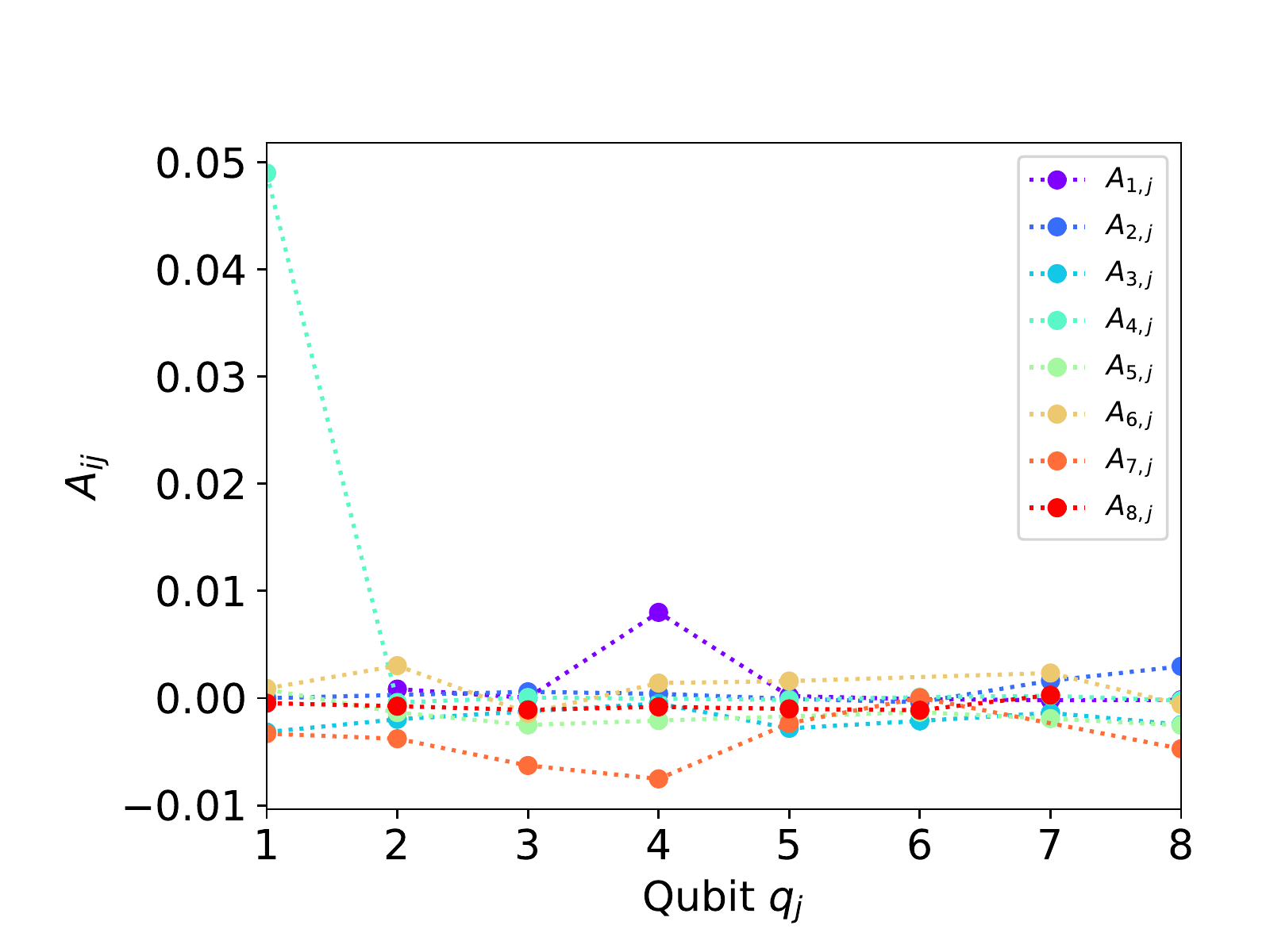} 
\caption{$A_{ij}$ for the 8-qubit register $C_8$. Each circuit was measured with 32k samples. Here $j=1,\cdots , 8$ refers to the qubit position along the chain $Q_{14}, Q_{13}, Q_{12}, Q_{11}, Q_{10}, Q_{9}, Q_{8}, Q_{7}$. In this register $A_{41}=0.049$.}
\label{Acorrelator8}
\end{figure}

The $A$ and $B$ correlators measure the sensitivity of expectation values to the states of spectator qubits:
\begin{eqnarray}
A_{ij} &=&  \langle E_{0}^{(i)} \rangle_{0 \cdots 0} - 
\langle E_{0}^{(i)} \rangle_{{\rm NOT}_j(0 \cdots 0) } 
\label{defA}  \\
B_{ijl} &=& \langle E_0^{(i)} E_0^{(j)} \rangle_{0 \cdots 0} - \langle E_0^{(i)} E_0^{(j)} \rangle_{{\rm NOT}_l (0 \cdots 0)}  .
\label{defB}
\end{eqnarray}
Here $A_{ij}$ is the difference between expectation values of $E_0^{(i)}$ on initial states $|0\rangle^{\otimes n}$ and $\sigma_j^x |0\rangle^{\otimes n}$.
$B_{ijl}$ is the change in $\langle E_0^{(i)} E_0^{(j)} \rangle_{0^{\otimes n}}$ when the initial state of spectator qubit $l \neq i,j$ is flipped. In a condensed matter physics setting, $A$ and $B$ might be called response functions that measure the change of $\langle E_{0}^{(i)} \rangle_{0^{\otimes n}}$ and $ \langle E_0^{(i)} E_0^{(j)} \rangle_{0^{\otimes n}}$ to a nearby spin flip. Correlators similar to $A$ and $B$ but based on $E_{1}$ instead of $E_{0}$ can also be defined, but these are not independent of $A, B$. Correlators defined with respect to an arbitrary initial state $x' \neq 0^{\otimes n}$ are similar in magnitude to $A, B$ and are not needed here. The measured $A$ correlators for the two registers are shown in Figs.~\ref{Acorrelator4} and \ref{Acorrelator8}, and the largest (in magnitude) elements of $A$ and $B$ are summarized in Table \ref{correlatorSummaryTable}.

\begin{table}[htb]
\centering
\caption{Values of the largest (in magnitude) correlators measured on the two ibmq\_16\_melbourne registers. For the $n=4$ case $i$ and $j$ refer to the qubit position along the chain $Q_{14}, Q_{13}, Q_{12}, Q_{11}$. For the $n=8$ case they refer to the position along the chain $Q_{14}, Q_{13}, Q_{12}, Q_{11}, Q_{10}, Q_{9}, Q_{8}, Q_{7}$. Each circuit was measured with 32k samples.}
\begin{tabular}{|c|c|c|c|}
\hline
$n$ & max $A_{ij}$  & max $B_{ijk}$ &  max $C_{ij}(0^{\otimes n})$   \\
\hline 
4 & $A_{41}$ = 4.7e-2 & $B_{241}$ = 4.7e-2 & $C_{23}$ = 2.0e-4  \\
8 & $A_{41}$ = 4.9e-2 & $B_{461}$ = 4.9e-2 & $C_{35}$ = 1.9e-4  \\
\hline
\end{tabular}
\label{correlatorSummaryTable}
\end{table}

The $C$ correlator quantifies correlated fluctuations of the measurement operators through their covariance
\begin{equation}
C_{ij}(x') = \big\langle \delta E_{0}^{(i)} \, \delta E_{0}^{(j)} \big\rangle_{\! x'} .
\label{defC} 
\end{equation}
In contrast to $A$ and $B$, we define $C$ with respect to an arbitrary initial  state $x' \in \{ 0,1 \}^n$ (this will be needed below). On the ibmq\_16\_melbourne device we find that $C_{ij}(x')$ with different $x'$ are similar in magnitude, and that the magnitude of the $C$-type correlations are 10 to 100 times smaller than that of $A$ and $B$. The largest elements of $C_{ij}$  with $x'=0000$ and $00000000$ are summarized in Table \ref{correlatorSummaryTable}. The $C$ correlator will be used in Sec.~\ref{Scalable T matrix estimation} to 
account for the effects of pair correlations on the $T$ matrix.

\subsection{Correlation volumes}
\label{Correlation volumes}

A critical component of our technique is the filtering of the set of initial classical states $x'  \in \lbrace 0,1\rbrace^n$ that have to be prepared and measured, reducing the total number of measurements required from $2^n$ to $O(4^k n^2)$. Here $k$ is the number of qubits in a {\it correlation volume}, which depends on the processor type and layout, the noise range and strength, and the desired accuracy of the estimated $T$. Informally, the correlation volume for qubit $i$ is the set of spectator qubits $j\neq i$ with which there are significant correlations.

To be precise, there are three correlation lengths or areas, which we generically call ``volumes". Each volume characterizes the spatial range of an $A$, $B$, or $C$ correlator. These can be measured directly (as in Sec.~\ref{Multiqubit measurement error correlators}) to determine the associated correlation volumes, the largest of which determines $k$.  However in the small processor studied here, the $A$ and $B$ correlation volumes (which measure sensitivity to spectator qubits) cover the entire device. So in addition to measuring $k$, we treat it as a {\it parameter} determining the size of the Moore neighborhoods, and we evaluate the accuracy of the resulting $T$ matrix as a function of $k$. When the ${\cal N}_i$ are large enough to contain the $A$-type and $B$-type correlation volumes (the entire register in our case), accurate $T$ estimation is achieved.

We also make a simplification concerning the $C$-type correlation volumes:  When measuring $C_{ij}(x')$ we can in principle restrict the qubit pairs $i$ and $j$ to lie within a correlation volume and set $C_{ij}(x')=0$ when they don't. However there is little benefit from doing this unless $n \gg k$, which is not the case here. Therefore we measure (and include in $T$) the $C$ correlators between all pairs of qubits in the register. We also note that although the $C$-type correlations are very small in our data, and might be neglected, we don't expect this to always be the case.

\section{Scalable T matrix estimation}
\label{Scalable T matrix estimation}

Here we explain the technique of scalable $T$ estimation in detail. We begin by discussing the physical basis of the technique. Error mitigation based on the $T$ matrix (\ref{defT}) is not scalable because it does not make use of:
\begin{enumerate}

\item The tensor product structure of the physical system, i.e., the fact that the qubits are composed of distinct subsystems or devices, such as ions or superconducting circuits.

\item The expectation that qubits mainly interact pairwise, and that multiqubit correlations are dominated by pair correlations. 
 
\item The assumption that for large enough devices, multiqubit correlations, however strong, will be finite-ranged and should decay at large distances. 
\end{enumerate}

The technique combines two distinct components. This is required to obtain an accurate $T$ matrix, as there are two distinct type of correlations present, those measured by the $A$ and $B$ correlators, and those measured by the $C$ correlators. The first component is to expand (\ref{mapped}) in powers of $C$-type SPAM fluctuations. This leads to the approximate $T$ matrix
\begin{equation}
T_{\rm est} =  T_{\rm mean}  + T_{\rm pair},
\label{defTest} 
\end{equation}
with matrix elements
\begin{eqnarray}
T_{\rm mean}(x|x') &=& \langle E_{x_1}^{(1)} \rangle_{x'}
\langle E_{x_2}^{(2)} \rangle_{x'}  \cdots  \langle E_{x_n}^{(n)} \rangle_{x'} \label{TmeanNonscalble}
  \\
T_{\rm pair}(x|x')  &=& \sum_{i<j} \bigg[ \langle \delta E_{x_i}^{(i)} \, \delta E_{x_j}^{(j)} \rangle_{x'} 
\! \times \! \prod_{\ell \neq i , j} \langle E_{x_\ell}^{(\ell)} \rangle_{x'} \bigg]  . \ \ \ 
\label{TprodNonscalble}
\end{eqnarray}
It is important to note that $T_{\rm mean}$ is distinct from $T_{\rm prod}$, defined in (\ref{Tprod}). This is because $T_{\rm prod}$ is a tensor product of $2 \times 2$ matrices, which neglects {\it all} ($A$, $B$, and $C$) correlations. $T_{\rm mean}$ becomes a strict tensor product if the spectator qubits are set to a background value such as 0 or 1, or averaged over. However here we will treat the spectator qubits using a filtering protocol, which will account for $A$-type correlations. $T_{\rm pair}$ accounts for $B$-type and $C$-type correlations.

The second component is best understood in terms of the matrix elements (\ref{TmeanNonscalble}) and (\ref{TprodNonscalble}). For a given initial state $x' \in \{0,1\}^n$, evaluating (\ref{TmeanNonscalble}) requires the measurement of $\langle E_{x_i}^{(i)} \rangle_{x'}$ for each qubit, and 
 (\ref{TprodNonscalble}) requires $\langle \delta E_{x_i}^{(i)} \, \delta E_{x_j}^{(j)} \rangle_{x'} $ for each pair, for a total of $O(n^2)$ measurements. But this applies to each of the $2^n$ initial states! However, in the absence of $A$-type and $B$-type correlations we could assume
\begin{eqnarray}
\langle E_{x_i}^{(i)} \rangle_{x'} & = & \langle E_{x_i}^{(i)} \rangle_{0 \cdots 0x'_i 0 \cdots 0} \\
\langle \delta E_{x_i}^{(i)} \, \delta E_{x_j}^{(j)} \rangle_{x'} & = & \langle \delta E_{x_i}^{(i)} \, \delta E_{x_j}^{(j)} \rangle_{0 \cdots 0x'_i 0 \cdots 0x'_j 0 \cdots 0},
\label{spectatorDecoupling} 
\end{eqnarray}
where all spectator qubits are initialized to 0 (or some other convenient state). Then there would only be $O(n^2)$ distinct quantities to measure to evaluate the $T$ matrix, namely
$\langle E_{x_i}^{(i)} \rangle_{0 \cdots x'_i  \cdots 0}$ and
$ \langle \delta E_{x_i}^{(i)} \, \delta E_{x_j}^{(j)} \rangle_{0 \cdots 0x'_i  \cdots x'_j 0 \cdots 0}$ for all $x_i, x_j, x'_i, x'_j \in \{0,1\}$ and  $i,j =1,2, \cdots , n.$ 
Motivated by this observation, we approximate
\begin{eqnarray}
\langle E_{x_i}^{(i)} \rangle_{x'} & \approx & \langle E_{x_i}^{(i)} \rangle_{f_{i}(x')} \\
\langle \delta E_{x_i}^{(i)} \, \delta E_{x_j}^{(j)} \rangle_{x'} & \approx & \langle \delta E_{x_i}^{(i)} \, \delta E_{x_j}^{(j)} \rangle_{f_{ij}(x')} ,
\label{spectatorFiltering} 
\end{eqnarray}
where $f_{i}(x)$ and $f_{ij}(x)$ are filters (Sec.~\ref{Qubit filters}) that set the initial states of spectator qubits outside of the Moore neighborhoods ${\cal N}_i$  and ${\cal N}_i \cup {\cal N}_j$ to 0.

The protocol for scalable $T$ estimation with a given set of size-$k$ Moore neighborhoods is implemented as follows: 
\begin{enumerate}

\item Measure the quantities $\langle E_{x_i}^{(i)} \rangle_{f_{i}(x')}$ for each qubit $i$. This requires no more than $ 2 n 2^k$ distinct circuits, where $2 \! \times \! 2^k$ is the number of classical states $x'$ that need to be prepared for each qubit (two initial states $\{0,1\}$ of $x'_i$ and $2^k$ initial states $\{0,1\}^k$ of the spectators in ${\cal N}_i$), and there are $n$ qubits in the register. Qubit neighborhoods near the boundary of a register may have fewer than $k$ qubits, hence $ 2 n 2^k$ is an upper bound.

\item Measure the quantities $\langle \delta E_{x_i}^{(i)} \, \delta E_{x_j}^{(j)} \rangle_{f_{ij}(x')}$ for each pair $i,j$. This requires no more than 
\begin{equation}
2 n^2 4^k
\label{circuitCount}
\end{equation}
distinct circuits, because there are $4 \! \times \! 4^k$ classical states $x'$ that need to be prepared for each pair (four initial states of $x'_i, x'_j$ and $4^k$ initial states of the spectators in ${\cal N}_i \cup {\cal N}_j$), and there are $n(n-1)/2$ pairs in the register. Qubit neighborhoods near the boundary of a register may have fewer than $k$ qubits, and ${\cal N}_i \cup {\cal N}_j$ will have fewer than $2k$ qubits when ${\cal N}_i$ and ${\cal N}_j$ overlap, hence (\ref{circuitCount}) is an upper bound. Therefore, for a given neighborhood size $k$, the overall quantum complexity is $O(4^k n^2)$.

\item Then classically compute $T_{\rm est} = T_{\rm mean} + T_{\rm pair}$ element by element, using
\begin{equation}
T_{\rm mean}(x|x') = \prod_{i=1}^n \langle E_{x_i}^{(i)} \rangle_{f_{i}(x')} 
\label{defTmean}
\end{equation}
and
\begin{equation}
T_{\rm pair}(x|x') \! = \! \sum_{i<j} \! \bigg[ \! \langle \delta E_{x_i}^{(i)} \, \delta E_{x_j}^{(j)} \rangle_{f_{ij}(x')} 
\! \times \! \! \prod_{\ell \neq i , j} \langle E_{x_\ell}^{(\ell)} \rangle_{f_{l}(x')} \!
\bigg]  .
\label{defTpair} 
\end{equation}
Classical evaluation of $T_{\rm mean}$ is observed to have a single-core runtime of $t \! = \! O(n4^n)$. This is expected because each matrix element contains a product of $n$ factors, and there are $4^n$  elements.  $T_{\rm pair}$  has an empirical runtime $t \! = \! O(n^3 4^n)$ because there are $n(n-1)/2$ pairs. Hence the overall classical complexity is $O(n^3 4^n)$. 

\end{enumerate}

We apply this technique to the ibmq\_16\_melbourne registers $C_4$ and $C_8$. The resulting accuracy is summarized in Table \ref{registerSummaryTable}, where we give the error
\begin{equation}
 \| T_{\rm est} - T_{\rm meas} \|
\end{equation}
of $T_{\rm est}$ against the directly measured $T$ matrix. For the register $C_4$ we first give the accuracy for the case of $k\!=\!2$ Moore neighborhoods (the largest that can be accommodated in the register). While the complexity is $O(n^2)$, the accuracy is poor, reflecting the presence of large nonlocal $A$ and $B$ correlations. To confirm this we also consider the case where the neighborhood ${\cal N}_i$ contains {\it all} qubits in the register (other than $i$). In this case we have $k \! = \! O(n)$ and the resulting accuracy is high because $T_{\rm est}$ incorporates all type $A, B,$ and $C$ correlations. For the register $C_8$ we give the accuracy for the case of 6-qubit  neighborhoods  (the largest that can be accommodated) and for complete $k \! = \! O(n)$ neighborhoods. The $k=2,6$ examples in Table 
\ref{registerSummaryTable} demonstrate the polynomial complexity of our approach, while the $k=O(n)$ examples demonstrate its accuracy.

\begin{table}[htb]
\centering
\caption{Accuracy of scalable $T$ matrix estimation on ibmq\_16\_melbourne registers $C_4$ and $C_8$ \cite{dataTakingMarch4}. Here $k$ is the size of the Moore neighborhoods ${\cal N}_i$. The rows with $k=O(n)$ use neighborhoods containing the entire register. The matrix norms are defined in Sec.~\ref{Error measures}. Each circuit was measured with 32k samples.}
\begin{tabular}{|c|c|c|c|}
\hline
$n$ & $k$ &  $ \| T_{\rm est} - T_{\rm meas} \|_d $ & $ \| T_{\rm est} - T_{\rm meas} \|_{\rm max} $  \\
\hline 
4 & 2 & 4.5e-2 & 5.4e-2 \\
4 & $O(n)$ & 3.7e-4  & 3.3e-4 \\
\hline 
8 & 6 & 3.6e-2 & 5.6e-2 \\
8 & $O(n)$ & 1.5e-3 & 1.7e-3 \\
\hline 
\end{tabular}
\label{registerSummaryTable}
\end{table}

\section{Conclusions}
\label{Conclusions}

Motivated by a common SPAM error mitigation technique \cite{BialczakNatPhys10,NeeleyNat10,DewesPRL12,14114994,160304512,SongPRL17,181102292,190505720,180411326,TannuIEEE19,190411935,190503150,190708518,191000129,191001969,191113289,200104449,200601805}, we develop and apply an efficient method to characterize and correct multiqubit SPAM errors on a register of $n$ qubits. The technique assumes that correlated SPAM errors are dominated by pair correlations, and that, for large $n$, the range of the multiqubit measurement error correlations do not grow with system size. The number of distinct circuits that have to be measured is $O(4^k n^2)$, and the estimated $T$ matrix will be accurate whenever $k$ exceeds the number of qubits in the largest correlation volume. 

The efficient protocol does not provide a significant measurement savings unless $k \ll n$. Therefore we separately demonstrated the efficiency and accuracy of the technique, but were not able to demonstrate these attributes at the same time. In the future we hope to apply the technique to other quantum computing architectures and to registers that are larger than the correlation volumes. 

The $A$ and $B$-type correlations (measuring sensitivity to spectator qubits) were found to be much larger than the $C$-type correlations (measurement operator covariances) in the ibmq\_16\_melbourne chip. However we do not expect that this will always be the case. Therefore we have treated the $A, B,$ and $C$ correlators as equally important.

The main weakness of our technique and the principle roadblock preventing application to even larger registers is the {\it classical} processing used in the evaluation of $T_{\rm pair}$, which has complexity $O(n^3 4^n)$. We hope to address this limitation in the future.

It would also be interesting to study the SPAM error correlations over time. In particular, how do the results presented here, using data taken on March 4, 2020, change from calibration to calibration and from day to day? While a systematic study of SPAM error {\it drift} is beyond the scope of this work, we acquired a second complete data set on May 14, more than two months after the first, and found remarkably similar results.  The accuracy of $T_{\rm est}$ using the second data set, summarized in Table \ref{registerSummaryTableMay14} of Appendix \ref{Second data set}, is found to be very similar to the results given in Table \ref{registerSummaryTable}.

After completing this work we learned of a different approach to scalable SPAM correction based on cumulant expansions \cite{200601805}.

\begin{acknowledgments}

Data was taken using the {BQP} software package developed by the authors. We thank Robin Blume-Kohout, Ken Brown, Shantanu Debnath, Jay Gambetta, Alexander Korotkov, Benjamin Nachman, Matthew Neeley, and Erik Nielsen for their private communication. We're also grateful to IBM Research for making their devices available to the quantum computing community. This work does not reflect the views or opinions of IBM or any of their employees. 

\end{acknowledgments}

\appendix

\section{Frobenius error}

Here we show that
\begin{equation}
\lim_{n \rightarrow \infty} \lim_{\epsilon \rightarrow 0} \bigg \| \begin{pmatrix} 1-\epsilon & \epsilon \\ \epsilon & 1-\epsilon \\ \end{pmatrix}^{\! \otimes n}  - I \  \bigg \|_{\rm F} = n \, 2^{\frac{n}{2}} \epsilon ,
\label{frobeniusScaling}
\end{equation}
where $I$ is the $2^n \times 2^n$ identity. Let
\begin{equation}
\tau = \begin{pmatrix} 1-\epsilon & \epsilon \\ \epsilon & 1-\epsilon \\ \end{pmatrix}.
\end{equation}
Then $\tau^{\! \otimes n}$ for $n>1$ has the following properties:
\begin{enumerate}

\item Each column sums to 1.

\item The diagonal elements are equal to 
\begin{equation}
(1-\epsilon)^n = 1 - n \epsilon + O(\epsilon^2).
\end{equation}

\item Each off-diagonal element is of the form
\begin{equation}
\epsilon^m (1-\epsilon)^{n-m} = \epsilon^m \big( 1 + O(\epsilon) \big),
\end{equation}
where $m$ is an integer satisfying $1 \le m \le n$.

\end{enumerate}
Therefore, in the $\epsilon \rightarrow 0$ limit, the diagonal elements of 
\begin{equation}
\tau^{\! \otimes n}-I
\label{tauMinusI}
\end{equation}
are equal to $-n \epsilon$, and the off-diagonal elements of (\ref{tauMinusI}) must take values from the set
\begin{equation}
\big\lbrace \epsilon,  \epsilon^2, \cdots ,  \epsilon^n \big\rbrace.
\end{equation}
Furthermore, by property 1, in each column there must be exactly $n$ off-diagonal elements with value $\epsilon$, and these will dominate the Frobenius norm of (\ref{tauMinusI}) in the $\epsilon \rightarrow 0$ limit. Therefore we have
\begin{equation}
\lim_{\epsilon \rightarrow 0}
\| \tau^{\! \otimes n}-I \|_{\rm F}^2 =  [ n^2 \epsilon^2 + n \epsilon^2] 2^n ,
\end{equation}
where the quantity in square brackets is the contribution to the norm {\it squared} from one column, with the diagonal and off-diagonal contributions given separately, and $2^n$ is the number of columns. This leads to
\begin{equation}
 \lim_{\epsilon \rightarrow 0} \bigg \| \begin{pmatrix} 1-\epsilon & \epsilon \\ \epsilon & 1-\epsilon \\ \end{pmatrix}^{\! \otimes n}  - I \  \bigg \|_{\rm F} = \sqrt{n(n+1)} \, 2^{\frac{n}{2}} \epsilon ,
 \label{frobeniusScalingPrecise}
\end{equation}
for $n>1$. Using (\ref{frobeniusScalingPrecise}) we obtain the result (\ref{frobeniusScaling}), as required.

\begin{widetext}

\section{Single-qubit T matrices}
\label{Single-qubit T matrices}

The single-qubit $T$ matrices for the registers $C_4$ and $C_8$ are given here in Tables \ref{tableSingleQubitRegistersize4} and \ref{tableSingleQubitRegistersize8}. The SPAM errors given in the columns labelled ``$\epsilon \, ({\rm meas})$" are measured with the spectator qubits initialized to the 0 state. The columns labelled ``$\epsilon \, ({\rm IBM \ Q})$" are the SPAM error values reported by IBM's calibration data.

\begin{table}[htb]
\centering
\caption{Single-qubit $T$ matrices measured on ibmq\_16\_melbourne qubits $\{ Q_{14}, Q_{13}, Q_{12}, Q_{11} \}$ \cite{dataTakingMarch4}. Each circuit was measured with 32k samples.}
\begin{tabular}{|c|c|c|c|c|c|}
\hline
melbourne qubit & $T_i \ ({\rm spectators} \! = \!  |0\rangle)$ & $T_i \ ({\rm spectators} \! = \!  |1\rangle)$ & $T_i \ ({\rm ave} \! :  \,  k \! = \! 2) $ &  $\epsilon$ (meas) &  $\epsilon$ (IBM Q) \\
\hline 
$Q_{14}$   
& $ \begin{pmatrix} 0.996 & 0.099 \\ 0.004 & 0.901 \\ \end{pmatrix}$    
& $ \begin{pmatrix} 0.988 & 0.115 \\ 0.012 & 0.885 \\ \end{pmatrix}$ 
& $\begin{pmatrix} 0.997 & 0.100 \\ 0.003 & 0.900 \\ \end{pmatrix}$  
&  $5.1 \%$  
&  $4.0 \% $ 
\\
\hline 
$Q_{13}$
& $ \begin{pmatrix} 0.940 & 0.125 \\ 0.060 & 0.875 \\ \end{pmatrix}$    
& $ \begin{pmatrix} 0.938 & 0.115 \\ 0.062 & 0.885 \\ \end{pmatrix}$ 
& $\begin{pmatrix} 0.938 & 0.120 \\ 0.062 & 0.880 \\ \end{pmatrix} $  
&  $9.3\%$ 
&  $10.4\%$ 
\\
\hline
$Q_{12}$ 
& $\begin{pmatrix} 0.986 & 0.051 \\ 0.014 & 0.949 \\ \end{pmatrix}$    
& $\begin{pmatrix} 0.988 & 0.054 \\ 0.012 & 0.946 \\ \end{pmatrix}$ 
&  $\begin{pmatrix} 0.987 & 0.052 \\ 0.013 & 0.948 \\ \end{pmatrix}$  
&  $3.3\%$  
&  $4.6\%$ 
\\
\hline
$Q_{11}$
& $\begin{pmatrix} 0.999 & 0.063 \\ 0.001 & 0.937 \\ \end{pmatrix}$    
& $ \begin{pmatrix} 0.950 & 0.118 \\ 0.050 & 0.882 \\ \end{pmatrix}$ 
&  $\begin{pmatrix} 0.998 & 0.062 \\ 0.002 & 0.938 \\ \end{pmatrix} $  
&  $3.2\%$  
&  $3.1\%$ 
\\
\hline
\end{tabular}
\label{tableSingleQubitRegistersize4}
\end{table}

\begin{table}[htb]
\centering
\caption{Single-qubit $T$ matrices measured on ibmq\_16\_melbourne qubits $\{ Q_{14}, Q_{13}, Q_{12}, Q_{11}, Q_{10}, Q_{9}, Q_{8}, Q_{7} \}$ \cite{dataTakingMarch4}. Each circuit was measured with 32k samples.}
\begin{tabular}{|c|c|c|c|c|c|c|c|}
\hline
melbourne qubit & $T_i \ ({\rm spectators} \! = \!  |0\rangle)$ & $T_i \ ({\rm spectators} \! = \!  |1\rangle)$ & $T_i \ ({\rm ave} \! :  \,  k \! = \! 2) $ & $T_i \ ({\rm ave} \! :  \,  k \! = \! 4) $ & $T_i \ ({\rm ave} \! :  \,  k \! = \! 6) $ &  $\epsilon$ (meas) &  $\epsilon$ (IBM Q) \\
\hline 
$Q_{14}$   
& $ \begin{pmatrix} 0.998 & 0.097 \\ 0.002 & 0.903 \\ \end{pmatrix} $    
& $ \begin{pmatrix} 0.989 & 0.110 \\ 0.011 & 0.890 \\ \end{pmatrix} $ 
& $\begin{pmatrix} 0.997 & 0.100 \\ 0.003 & 0.900 \\ \end{pmatrix} $  
& $\begin{pmatrix} 0.994 & 0.103 \\ 0.006 & 0.897 \\ \end{pmatrix}$  
& $\begin{pmatrix} 0.994 & 0.104 \\ 0.006 & 0.896 \\ \end{pmatrix}$  
&  $5.0\%$  
&  $4.0\%$ 
\\
\hline 
$Q_{13}$
& $ \begin{pmatrix} 0.940 & 0.130 \\ 0.060 & 0.870 \\ \end{pmatrix} $    
& $ \begin{pmatrix} 0.923 & 0.128 \\ 0.077 & 0.872 \\ \end{pmatrix} $ 
& $\begin{pmatrix} 0.938 & 0.144 \\ 0.062 & 0.856 \\ \end{pmatrix} $  
& $\begin{pmatrix} 0.939 & 0.138 \\ 0.061 & 0.862 \\ \end{pmatrix}$  
& $\begin{pmatrix} 0.939 & 0.136 \\ 0.061 & 0.864 \\ \end{pmatrix}$  
&  $9.5\%$ 
&  $10.4\%$ 
\\
\hline
$Q_{12}$ 
& $\begin{pmatrix} 0.988 & 0.054 \\ 0.012 & 0.946 \\ \end{pmatrix}$    
& $\begin{pmatrix} 0.990 & 0.049 \\ 0.010 & 0.951 \\ \end{pmatrix} $ 
&  $\begin{pmatrix} 0.989 & 0.052 \\ 0.011 & 0.948 \\ \end{pmatrix} $  
& $\begin{pmatrix} 0.989 & 0.051 \\ 0.011 & 0.949 \\ \end{pmatrix}$  
& $\begin{pmatrix} 0.990 & 0.052 \\ 0.010 & 0.948 \\ \end{pmatrix}$  
&  $3.3\%$  
&  $4.6\%$ 
\\
\hline
$Q_{11}$
& $\begin{pmatrix} 0.999 & 0.061 \\ 0.001 & 0.939 \\ \end{pmatrix} $    
& $ \begin{pmatrix} 0.946 & 0.127 \\ 0.054 & 0.873 \\ \end{pmatrix}$ 
& $\begin{pmatrix} 0.999 & 0.061 \\ 0.001 & 0.939 \\ \end{pmatrix} $  
& $\begin{pmatrix} 0.999 & 0.062 \\ 0.001 & 0.938 \\ \end{pmatrix}$  
& $\begin{pmatrix} 0.975 & 0.091 \\ 0.025 & 0.909 \\ \end{pmatrix}$  
&  $3.1\%$  
&  $3.1\%$ 
\\
\hline 
$Q_{10}$
& $ \begin{pmatrix} 0.970 & 0.060 \\ 0.030 & 0.940 \\ \end{pmatrix} $    
& $\begin{pmatrix} 0.997 & 0.053 \\ 0.003 & 0.947 \\ \end{pmatrix} $ 
& $\begin{pmatrix} 0.977 & 0.059 \\ 0.023 & 0.941 \\ \end{pmatrix} $  
& $\begin{pmatrix} 0.977 & 0.058 \\ 0.023 & 0.942 \\ \end{pmatrix}$  
& $\begin{pmatrix} 0.987 & 0.056 \\ 0.013 & 0.944 \\ \end{pmatrix}$  
&  $4.5\%$ 
&  $4.0\%$ 
\\
\hline
$Q_{9}$ 
& $\begin{pmatrix} 0.987 & 0.080 \\ 0.013 & 0.920 \\ \end{pmatrix} $    
& $\begin{pmatrix} 0.982 & 0.087 \\ 0.018 & 0.913 \\ \end{pmatrix} $ 
& $\begin{pmatrix} 0.986 & 0.081 \\ 0.014 & 0.919 \\ \end{pmatrix}$  
& $\begin{pmatrix} 0.986 & 0.079 \\ 0.014 & 0.921 \\ \end{pmatrix}$  
& $\begin{pmatrix} 0.985 & 0.084 \\ 0.015 & 0.916 \\ \end{pmatrix}$  
&  $4.7\%$  
&  $4.8\%$ 
\\
\hline
$Q_{8}$
& $\begin{pmatrix} 0.692 & 0.329 \\ 0.308 & 0.671 \\ \end{pmatrix} $    
& $ \begin{pmatrix} 0.736 & 0.297 \\ 0.264 & 0.703 \\ \end{pmatrix} $ 
&  $\begin{pmatrix} 0.695 & 0.329 \\ 0.305 & 0.671 \\ \end{pmatrix} $  
& $\begin{pmatrix} 0.697 & 0.328 \\ 0.303 & 0.672 \\ \end{pmatrix}$  
& $\begin{pmatrix} 0.717 & 0.304 \\ 0.283 & 0.696 \\ \end{pmatrix}$  
&  $31.8\%$  
&  $27.1\%$ 
\\
\hline
$Q_{7}$
& $\begin{pmatrix} 0.997 & 0.131 \\ 0.003 & 0.869 \\ \end{pmatrix} $    
& $ \begin{pmatrix} 0.996 & 0.153 \\ 0.004 & 0.847 \\ \end{pmatrix} $ 
& $\begin{pmatrix} 0.997 & 0.130 \\ 0.003 & 0.870 \\ \end{pmatrix}$  
& $\begin{pmatrix} 0.997 & 0.135 \\ 0.003 & 0.865 \\ \end{pmatrix} $  
& $\begin{pmatrix} 0.997 & 0.113 \\ 0.003 & 0.887 \\ \end{pmatrix}$  
&  $6.7\%$  
&  $7.5\%$ 
\\
\hline
\end{tabular}
\label{tableSingleQubitRegistersize8}
\end{table}

\end{widetext}

\section{Second data set}
\label{Second data set}

A second complete data set was taken on May 14, 2020, using the same 
ibmq\_16\_melbourne registers $C_4$ and $C_8$. The results were found to be remarkably similar to the data presented above, which was taken on March 4, 2020. The accuracy of $T$ matrix estimation with the second data set is summarized in Table \ref{registerSummaryTableMay14}, which can be compared to Table \ref{registerSummaryTable}.

\begin{table}[htb]
\centering
\caption{Accuracy of scalable $T$ matrix estimation on ibmq\_16\_melbourne registers $C_4$ and $C_8$. Each circuit was measured with 32k samples. (This data was taken on May 14, 2020).}
\begin{tabular}{|c|c|c|c|}
\hline
$n$ & $k$ &  $ \| T_{\rm est} - T_{\rm meas} \|_d $ & $ \| T_{\rm est} - T_{\rm meas} \|_{\rm max} $  \\
\hline 
4 & 2 & 3.4e-2 & 3.8e-2 \\
4 & $O(n)$ & 4.2e-4  & 3.7e-4 \\
\hline 
8 & 6 & 3.4e-2 & 5.9e-2 \\
8 & $O(n)$ & 1.7e-3 & 1.6e-3 \\
\hline 
\end{tabular}
\label{registerSummaryTableMay14}
\end{table}

\clearpage

\bibliography{/Users/mgeller/Dropbox/bibliographies/algorithms,/Users/mgeller/Dropbox/bibliographies/applications,/Users/mgeller/Dropbox/bibliographies/books,/Users/mgeller/Dropbox/bibliographies/cm,/Users/mgeller/Dropbox/bibliographies/dwave,/Users/mgeller/Dropbox/bibliographies/control,/Users/mgeller/Dropbox/bibliographies/general,/Users/mgeller/Dropbox/bibliographies/group,/Users/mgeller/Dropbox/bibliographies/ions,/Users/mgeller/Dropbox/bibliographies/math,/Users/mgeller/Dropbox/bibliographies/ml,/Users/mgeller/Dropbox/bibliographies/nmr,/Users/mgeller/Dropbox/bibliographies/optics,/Users/mgeller/Dropbox/bibliographies/qec,/Users/mgeller/Dropbox/bibliographies/simulation,/Users/mgeller/Dropbox/bibliographies/software,/Users/mgeller/Dropbox/bibliographies/superconductors,/Users/mgeller/Dropbox/bibliographies/surfacecode,/Users/mgeller/Dropbox/bibliographies/tn,endnotes}

\begin{thebibliography}{23}
\bibitem [{moo()}]{mooreNote}%
  \BibitemOpen
  \href@noop {} {}\bibinfo {note} {Note that some authors define a Moore
  neighborhood ${\cal N}_i$ to include site $i$.}\BibitemShut {Stop}%
\end{thebibliography}

\end{document}